\documentclass[pre,twocolumn,amsmath,amssymb,floatfix,superscriptaddress]{revtex4}

\usepackage{graphicx}
\usepackage{dcolumn}
\usepackage{bm}
\usepackage{subfigure}

\graphicspath{{images/}{plots/}}

\newcommand{\avg}[1]{\ensuremath{\left< #1 \right>}}

\newcommand{\brac}[1]{\ensuremath{\left(#1\right)}}

\DeclareMathOperator{\ee}{e}

\begin{document}

    \title{Large deviations of the length of the longest increasing subsequence of random permutations and random walks}
    \author{J\"orn B\"orjes}
    \email{joern.boerjes@uni-oldenburg.de}
    \author{Hendrik Schawe}
    \email{hendrik.schawe@uni-oldenburg.de}
    \author{Alexander K. Hartmann}
    \email{a.hartmann@uni-oldenburg.de}
    \affiliation{Institut f\"ur Physik, Universit\"at Oldenburg, 26111 Oldenburg, Germany}
    \date{\today}

    \begin{abstract}
        We study numerically the distributions of the length $L$ of the longest
        increasing subsequence (LIS) for the two cases of random permutations
        and of one-dimensional random walks.
        Using sophisticated large-deviation algorithms, we are able to
        obtain very large parts of the distribution, especially also covering
        probabilities smaller than $P(L) = 10^{-1000}$.
        This enables us to verify for the length of the LIS
        of random permutations the analytically
        known asymptotics of the rate function and
        even the whole Tracy-Widom distribution, to which we observe
        a rather fast convergence in the larger than typical part.
        For the length $L$ of LIS of random walks, where
        no analytical results are known to us, we test a proposed scaling
        law and observe convergence of the tails into a collapse for increasing
        system size. Further, we obtain estimates for the leading order behavior
        of the rate functions of both tails.
    \end{abstract}


    \maketitle

\section{Introduction}
    We study the distribution of the length $L$ of the \emph{longest
    increasing subsequence} (LIS) \cite{romik2015}
    of different ensembles of random sequences.
    Here, a subsequence of a given sequence is obtained by removing
    arbitrary entries and keeping the order of the remaining entries.
    In particular, the remaining entries are not necessarily
    neighbors in the given sequence. For a LIS it is required
    that the remaining entries are increasing from left
    to right and the number of remaining elements is maximal.
    An application of the LIS is for aligning whole genomes \cite{delcher1999}.
    The first mention of this problem seems to be from Stanis\l{}aw Ulam
    \cite{beckenbach2013modern},
    and is therefore also known as ``Ulam's problem''. In his study the
    mean length $L$ of LIS on random permutations (RP) of $n$ integers were
    scrutinized by means of Monte Carlo simulations and it was conjectured that in the
    limit of large $n$, the length converges to $L = c \sqrt{n}$, with some constant $c$,
    which was later proven to be $c=2$ \cite{aldous1999longest}. In the following years much work
    was published scrutinizing the large deviation behavior of this problem and
    explicit expressions for both the left (lower) and right (upper) tail
    were derived rigorously \cite{Seppalainen1998,logan1977variational,deuschel1999increasing}.
    Interestingly, for the LIS of the random permutation it was shown that the
    distribution $P(L)$ of its length is a Tracy-Widom distribution \cite{baik1999distribution}.
    The Tracy-Widom distribution was at that time only known from random matrix theory,
    where it described the distribution of the largest eigenvalues of the
    \emph{Gaussian unitary ensemble} (GUE), an ensemble of Hermitian random
    matrices. In physics it came into focus after an explicit mapping of a
    $1+1$ dimensional polynuclear growth model \cite{Prahofer2000universal}.
    Subsequently other mappings of $1+1$ dimensional growth models belonging to
    the Kardar-Parisi-Zhang universality like an anisotropic ballistic
    deposition \cite{majumdar2004anisotropic} were found. Other models in which
    the Tracy-Widom distribution appears, include the totally asymmetric
    exclusion process \cite{Johansson2000} and directed polymers \cite{Baik2000}.
    For a pedagogical overview about the relations of different models exhibiting
    a Tracy-Widom distribution, we recommend Ref.~\cite{bouchaud2011complex}.
    Fluctuations in growth processes following the Tracy-Widom distribution
    could also be observed in experiments, e.g., from growing liquid crystals
    where the Tracy-Widom distribution of the GUE appears for circular growth
    and of the \emph{Gaussian orthogonal ensemble} (GOE) for growth from a flat
    surface \cite{takeuchi2010universal,takeuchi2011growing}.

    The Tracy-Widom distribution seems to occur always together with a
    \emph{third order phase transition} between a
    \emph{strongly-interacting} phase in the left tail and a \emph{weakly-interacting}
    phase in the right tail, whose crossover is characterized by the Tracy-Widom
    distribution \cite{Majumdar2014Top}. For these third order phase transitions,
    the probability density function behaves in the left tail as $P(x) \approx \ee^{-n\Phi_{-}}$
    with the role of the free energy played by the \emph{rate function}
    $\Phi_{-}(x) \sim (a-x)^3$ for $x \to a$ from the left, where $a$ is the critical point
    of the transition, i.e., the scaled mean value. Here, $n$ is some large
    parameter, e.g, the system size. The $O(x^3)$ leading order behavior of
    $\Phi_-$ generally leads to a discontinuity in the third derivative of the free energy
    and therefore to a third order phase transition. This seems to be a characteristic sign
    predicting the main region of the distribution to follow a Tracy-Widom distribution.
    Therefore the behavior of the far tails of problems of this universality
    are of great interest to understand this connection better. Consequently
    the \emph{large deviations} of some of these models were studied thoroughly
    \cite{Majumdar2014Top, Doussal2016Large}.

    For the distribution of the length of the LIS of random permutations there
    are also analytical results for the large deviations, i.e., the
    behavior for large values of $n$ including the far tails
    \cite{Seppalainen1998,logan1977variational,deuschel1999increasing,baik1999distribution},
    which also show the characteristic behavior of the above mentioned left-tail
    rate function.
    For the case of the length of the LIS of random walks, bounds for the
    behavior of the mean are known \cite{angel2017increasing} and  there is also
    numerical work which is concerned with the distribution in the
    typical region \cite{mendoncca2017empirical}, i.e., those LIS
    which occur with a high enough probability of about $\ge 10^{-6}$. We deem
    it worthwhile to look also for this system closer at the tails of the
    distribution for finite systems.

    For the purpose of studying the large deviations of this problem numerically,
    we will utilize sophisticated large deviation sampling methods
    to observe the distribution of the length $L$ for two ensembles of random
    sequences. This way we can observe directly the far tails of the
    Tracy-Widom distribution for the random permutation case \cite{baik1999distribution} and can confirm
    the known large $n$ asymptotics \cite{deuschel1999increasing}. The second ensemble are
    one-dimensional random walks with increments from a uniform distribution.
    While we can observe the scaling proposed in Ref.~\cite{mendoncca2017empirical}
    for the main region, the tails are subject to considerable finite-size effects.
    Nevertheless the distributions collapse over larger regions for larger sizes $n$.
    Also, we give estimates for the leading order behavior of the rate functions
    governing the left and right tail of the distribution $P(L)$.

    This study will first introduce the different ensembles of interest and
    the algorithms used to obtain the distribution of the length
    in Sec.~\ref{sec:methods}. In Sec.~\ref{sec:results} we will show the
    results we gathered and interpret them. We conclude this study in
    Sec.~\ref{sec:conclusion}.

\section{Models and Methods}
\label{sec:methods}
    To define the longest increasing subsequence (LIS), we have to define a
    subsequence first. Given some sequence $S=(S_1,S_2,\ldots,S_n)$
    a \emph{subsequence} of length $L$ is a sequence $s=(S_{i_1},S_{i_2}, \ldots, S_{i_L})$
    ($1\le i_j \le n, i_j<i_{j+1}$ for all $j=1,\ldots,L$)
    containing only elements present in $S$ in the same order as in $S$, though
    possibly with gaps. An \emph{increasing subsequence} has elements such that every
    element in $s$ is smaller than its predecessor,
    i.e., $S_{i_j} < S_{i_{j+1}}$ for $j=1,\ldots,L-1$. The LIS is consequently the
    longest, i.e., the one with the highest number $L$ of
    elements, of all possible increasing
    subsequences. Note that the LIS is not uniquely defined, but by definition
    its length is unique. As an example two different LIS are marked by overlines and underlines
    in the following sequence: $S = (\underline{3}, 9, \underline{4}, \overline{1}, \overline{2}, \underline{7}, \overline{6}, \overline{\underline{8}}, 0, 5)$

\begin{figure}[htb]
        \centering
        \subfigure[\label{fig:ex:rp} random permutation]{
            \includegraphics[scale=1]{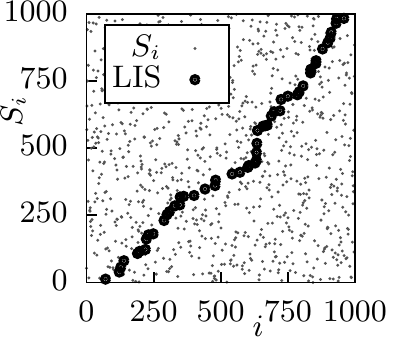}
        }
        \subfigure[\label{fig:ex:rw} random walk]{
            \includegraphics[scale=1]{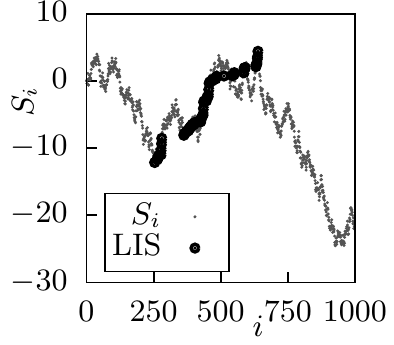}
        }
        \caption{\label{fig:ex}
            Visualization of random sequences of length $n=1000$ where the
            value is plotted over the corresponding index. Marked with circles
            are the entries of one possible LIS. \subref{fig:ex:rp} random
            permutation, \subref{fig:ex:rw} random walk.
        }
    \end{figure}

    In this study the sequence $S$ is either drawn from the ensemble of \emph{random
    permutations} of $n$ consecutive integers or from the ensemble of \emph{random
    walks} with increments $\delta_j$ $(j=1,\ldots,n)$ from a uniform
    distribution $\delta_j \sim U(-1,1)$, such that
    \begin{align}
        \label{eq:rw}
        S_i = \sum_{j=1}^i \delta_j.
    \end{align}
    An example of each sequence with the corresponding LIS marked is shown in
    Fig.~\ref{fig:ex}. Here the typical difference between the random permutation
    and random walks are visible: The entries of the random walk are strongly
    correlated such that the random walk typically consists of runs with
    downward or upward trends, such that the LIS is typically confined in an
    upward trend and its entries therefore are close together. The random
    permutation, on the other hand, typically shows LIS with entries over the
    whole range. Therefore it is plausible that the distributions of the length
    of the LIS for these two ensembles differ \cite{mendoncca2017empirical}.

    To find the LIS of any given sequence, we use the
    \emph{patience sort algorithm}, which is originally a sorting algorithm.
    This choice is mainly motivated by the simplicity of the algorithm if one
    is only interested in the length of the LIS. We will only introduce the
    very simple version to obtain the length, but a comprehensive review of the
    connection of patience sort with the LIS can be found in
    Ref.~\cite{aldous1999longest}. In short, the patience sort algorithm works
    as follows:
    We iterate over the $n$ entries $S_i$ and place each into an initially
    empty stack (or pile) $a_j$
    on the smallest $j$ such that for the top entry $\mathrm{top}(a_j) > S_i$
    holds. Note that this will always ensure
    that the top entries of $a$ are ascendingly sorted, such that we can
    determine $j$ by a binary search in $\mathcal{O}(\ln n)$. Finally, the
    number of non-empty stacks $a_j$ is equal to
    the length $L$ of the LIS.

    \subsection{Large-deviation Sampling}
    To be able to gather statistics of the large-deviation regime numerically
    \cite{practical_guide2015},
    we need to apply a sophisticated sampling scheme. Therefore we use a well
    tested \cite{Hartmann2002Sampling,Hartmann2011,Hartmann2014high} Markov
    chain Monte Carlo sampling which treats the system as a physical system at
    some artificial \emph{temperature} with the observable of interest as
    its \emph{energy}. Since the algorithm has been presented
    comprehensively in the literature, we here only state the details specific
    to the current application.
    In our case, we identify the state of the system with the
    sequence, the length $L$ with the energy and
    sample the equilibrium state at temperature $\Theta$ using the
    Metropolis algorithm \cite{metropolis1953equation,newman1999monte}.
    Controlling the temperature allows us to direct the sampling to different
    regimes of the distributions, to eventually cover the distributions
    over a large part of the support.
    To evolve our Markov chain of sequences, we have to
    introduce change moves, which modify a sequence and consequently the energy
    $L$. For the random permutation we swap two random entries and for the
    random walk we replace one of the increments $\delta_j$ (cf.~Eq.~\eqref{eq:rw})
    by a new random number drawn from the same uniform distribution.
    These changes are accepted according to the Metropolis acceptance ratio
    \begin{align}
        \label{eq:accept}
        P_{\mathrm{acc}} = \min(1,\ee^{-\Delta L/\Theta}),
    \end{align}
    where $\Delta L$ is the change in energy due to the change move. This
    Markov chain of sequence realizations will converge to an equilibrium
    state. As usual with Markov
    chain Monte Carlo simulations, we need to ensure equilibration and
    that the samples are decorrelated \cite{newman1999monte}.

\begin{figure}
        \includegraphics[scale=1]{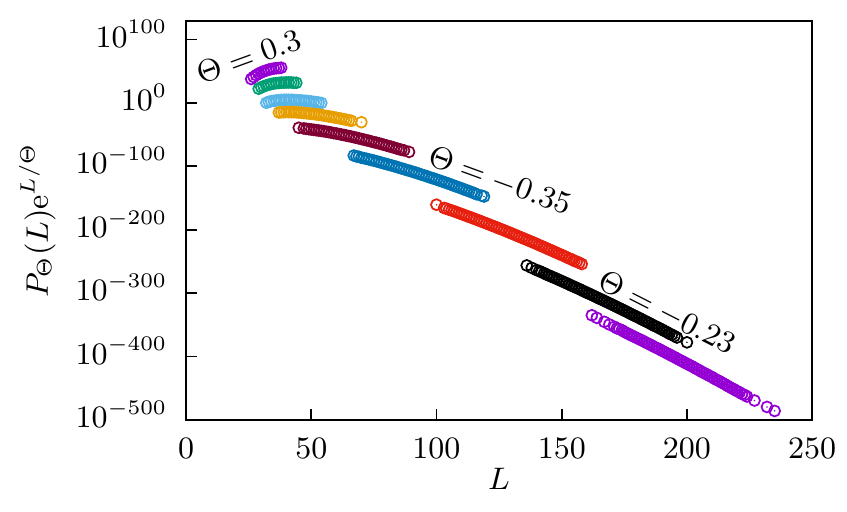}
        \caption{\label{fig:shifted}
            Intermediate step after correction with Eq.~\ref{eq:correction}
            but before determination of the values $Z_{\Theta_i}$ (i.e., all $Z_{\Theta_i} = 1$).
            The data is gathered for random walk sequences of length $n=512$.
            Each shade of gray (color) is sampled at a different temperature $\Theta$,
            for three datasets the corresponding temperatures are annotated.
            (For clarity some evaluated temperatures are omitted.)
        }
    \end{figure}

    As should be intuitively plausible, in equilibrium the realizations will
    generally have a lower than typical energy for low temperatures and typical
    energies for high temperatures. We can also introduce negative temperatures
    for larger than typical energies. This way the temperature can be tuned to
    guide the simulation towards realizations within a specific range of
    energies $L$. Since we know the equilibrium distribution $Q_\Theta(S)$
    at temperature $\Theta$ of realizations, i.e., sequences $S$, to be
    \begin{align}
        Q_\Theta(S) = \frac{1}{Z_\Theta} \ee^{-L(S)/\Theta} Q(S),
    \end{align}
    with the natural distribution $Q(S)$,
    we can later correct for the bias introduced by the temperature and arrive
    at the unbiased distribution $P(L)$ with good statistics also in the regions
    unreachable by simple sampling. Therefore consider the sampled equilibrium
    distributions $P_\Theta(L)$. To connect them to the distribution of realizations $Q_\Theta(S)$,
    we can sum all realizations with the same value of $L$, leading to
    \begin{align}
        P_\Theta(L) &= \sum_{\{S | L(S) = L\}} Q_\Theta(S)\\
            &= \sum_{\{S | L(S) = L\}} \frac{1}{Z_\Theta} \ee^{-L(S)/\Theta} Q(S)\\
            \label{eq:correction}
            &= \frac{1}{Z_\Theta} \ee^{-L(S)/\Theta} P(L).
    \end{align}
    Solving this equation for $P(L)$ allows to correct for the bias introduced
    by the temperature. An intermediate snapshot of this process is shown in
    Fig.~\ref{fig:shifted}.

    The constants $Z_\Theta$ can be obtained by enforcing continuity of the
    distribution, i.e.,
    \begin{align}
        P_{\Theta_j}(L) \ee^{L/\Theta_j} Z_{\Theta_j} = P_{\Theta_i}(L) \ee^{L/\Theta_i} Z_{\Theta_i}
    \end{align}
    for pairs of $i$, $j$ for which the gathered data $P_{\Theta_i}(L)$ overlaps
    with $P_{\Theta_j}(L)$. While this can be used to approximate the ratios of
    pairwise $Z_{\Theta_i}$, the absolute value can then be obtained by normalization of
    the whole distribution. This procedure requires a clever choice of
    temperatures, since gaps in the sampled range of $L$ would make it
    impossible to find a ratio of $Z_{\Theta_i}$ on the left and right side of
    the gap. We used in the order of $100$ distinct temperatures. In general,
    the larger the size $n$, the more temperatures are needed.

\section{Results}
\label{sec:results}
    We applied the temperature-based sampling scheme to obtain the probability
    distributions of the length of the LIS for the two cases
    of random permutations and of random
    walks with uniform increments. In both cases, we studied  five
    different system sizes $n$ up to $n=4096$ each.

\subsection{Random Permutations}
    First, we will look at the distribution of the length of the LIS of random
    permutations. For this case there are already a lot of properties known
    in the limit of $n\to\infty$.

\begin{figure}
        \includegraphics[scale=1]{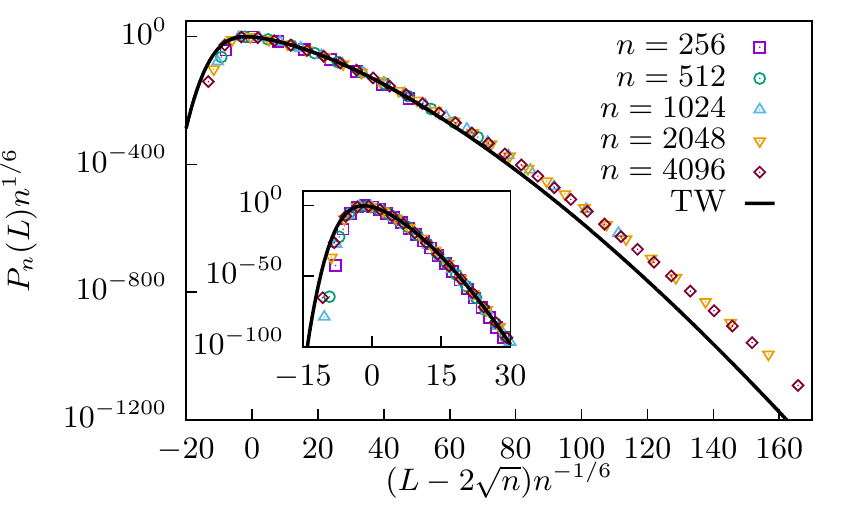}
        \caption{\label{fig:finite_rp}
            Numerically obtained distributions for different system
            sizes $n$ rescaled according to Eq.~\eqref{eq:tw}. The Tracy-Widom
            distribution is drawn as a black line \cite{Prahofer2004} and is expected
            to be the curve all distributions collapse onto. The inset shows
            a zoom on the intermediate tails.
            On the left the tendency of our data towards the Tracy-Widom
            distribution with increasing system size $n$ is visible.
            (For clarity some datapoints are discarded to show the same density
            of symbols for every system size.)
        }
    \end{figure}

    It is known that the distribution should converge to a
    suitably rescaled Tracy-Widom distribution $\chi$ of the GUE ensemble
    \cite{baik1999distribution} for large values of $n$ as
    \begin{align}
        \label{eq:tw}
        P_n\brac{\brac{L - 2\sqrt{n}}n^{-1/6}} = \chi\brac{\brac{L - 2\sqrt{n}}n^{-1/6}}.
    \end{align}
    Rescaled to accompensate this leading behavior, our
    results are shown in Fig.~\ref{fig:finite_rp}. By using the
    large-deviation approach, we are able to measure probabilities as small
    as $10^{-1000}$ and below, allowing us to go beyond the first numerical work
    \cite{mendoncca2017empirical} on the distribution of LIS.
    We can observe a very good collapse up to probabilities of $10^{-200}$ of
    our data onto the Tracy-Widom distribution given in the tables of
    Ref.~\cite{Prahofer2004}.

    Also note that the collapse works very well in the intermediate right tail
    but converges a bit slower in the left tail and far slower in the far-right
    tail. The inset zooms into the intermediate tail
    of the probability density function $P > 10^{-100}$, where the collapse
    fits very well to the expected Tracy-Widom distribution.
    In the far tails we observe considerable deviations,
    from the tabulated data, which are at least in part caused by finite-size
    effects due to the relatively small sizes $n$ of our sequences. For a more
    extensive study of these finite-size effects, one could obtain the empirical
    distribution for more sizes, and extrapolate the finite-size effects to
    $n\to\infty$, as done in \cite{schawe2018ground}. Nevertheless, our
    numerically obtained tails fit very well to another expected form which will
    be explained later, such that we assume a stronger influence of finite-size
    effects in the far tails for this scaling instead of systematic errors.

    Also note that while we can sample a very large part of the distribution
    of the length of longest subsequences of random permutations even including
    events with a probability less than $10^{-1000}$ for the largest
    permutations, we can not reach across the whole range of possible values
    and would possibly need to modify our sampling algorithm by either
    switching to a better change move or consider a different sampling like
    Wang-Landau's method \cite{Wang2001Efficient}.

    The left tail asymptotic, i.e., $L/\sqrt{n} = x < 2$, of the probability
    density function is given by the analytically known rate function \cite{logan1977variational,deuschel1999increasing}
    \begin{align}
        \label{eq:H}
        \lim_{n\to\infty} \frac{1}{n} \ln P_n(L) = -2H_0(x)
    \end{align}
    with
    \begin{align}
        \nonumber
        H_0(x) =& -\frac{1}{2} + \frac{x^2}{8} + \ln\frac{x}{2} \\
               &- \brac{1+\frac{x^2}{4}} \ln \brac{\frac{2x^2}{4+x^2}};
    \end{align}
    the right tail asymptotic, i.e., $L/\sqrt{n} = x > 2$, is given by \cite{Seppalainen1998,deuschel1999increasing}
    \begin{align}
        \label{eq:U}
        \lim_{n\to\infty} \frac{1}{\sqrt{n}} \ln P_n\brac{L} = -U_0(x)
    \end{align}
    with
    \begin{align}
        U_0(x) = 2x\cosh^{-1}\brac{x/2} - 2\sqrt{x^2-4}.
    \end{align}
    Note that Eq.~\eqref{eq:U} behaves atypically for a rate function as the
    distribution behaves like $P_n\propto \ee^{-\sqrt{n}U_0}$, which according
    to the definition, e.g., given in \cite{Touchette2009large}, does therefore not
    fulfill the large deviation principle. Nevertheless, it describes
    the behavior of the distribution in leading order.

\begin{figure}
        \includegraphics[scale=1]{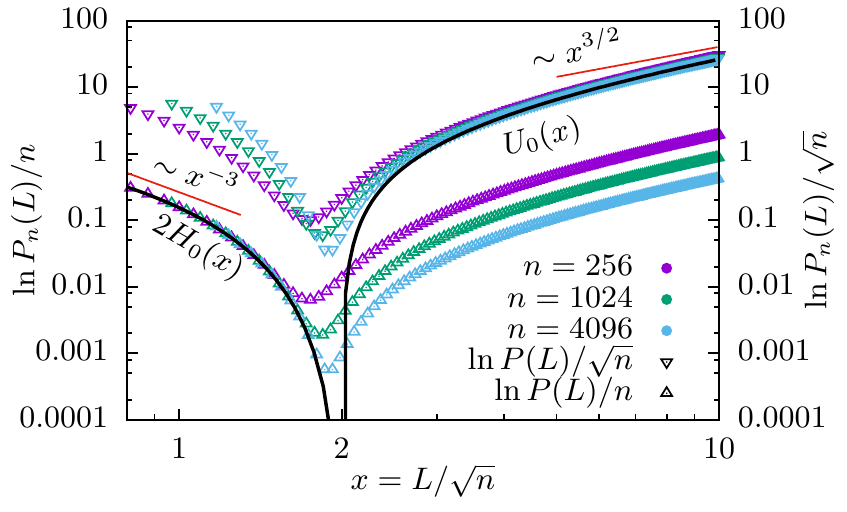}
        \caption{\label{fig:rate_rp}
            Empirical rate functions for different system sizes $n$. On the top
            (triangles down) scaled as $\ln P_n(L)/\sqrt{n}$ to emphasize the
            right tail behavior. On the bottom (triangles up)
            scaled as $\ln P_n(L)/n$ to emphasize the left
            tail behavior. The analytically known rate functions for both
            tails $2H_0$ and $U_0$ are shown in the correspondingly scaled region
            and a convergence of the data to these functions is well visible.
            The leading order terms of the series expansion (cf. \cite{Seppalainen1998})
            are also shown as straight lines next to the rate function.
        }
    \end{figure}

    We use our sampled data to test these rate functions. If the data are
    suitably rescaled according to Eq.~\eqref{eq:H} and \eqref{eq:U}, in the
    corresponding tails we can observe a very nice convergence of the data to
    the rate functions. This is plotted in Fig.~\ref{fig:rate_rp}.
    This excellent agreement of analytical and numerical
    results over hundreds of decades in probability, gives us confidence
    that our approach works well and can be extended to cases where
    no analytical results are known. Also note that we can observe in our data
    the leading order behavior of the left tail rate function $H_0$, which
    goes with the exponent $3$ characteristic for the third order
    phase transition confirming its connection with the Tracy-Widom
    distribution \cite{Majumdar2014Top}.

\subsection{Random Walks}
\begin{figure}
        \includegraphics[scale=1]{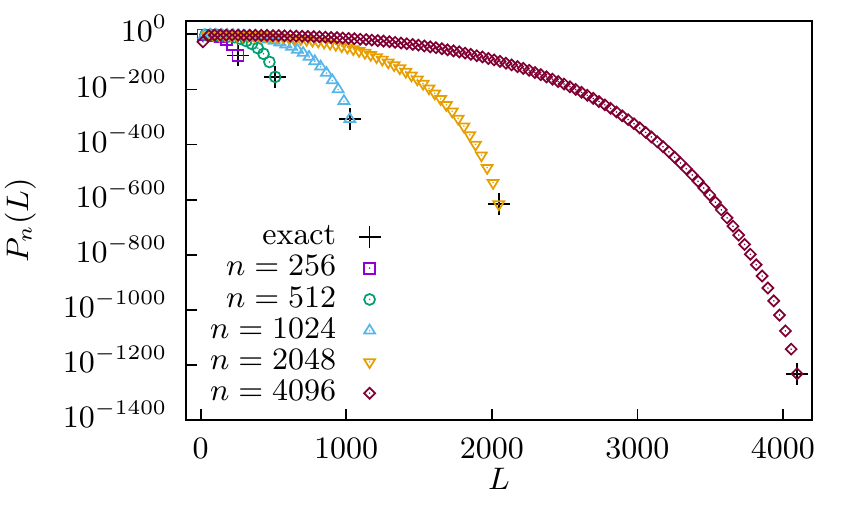}
        \caption{\label{fig:verteilungRW}
            Probability distributions $P_n(L)$ of the length of LIS of random
            walks with exact extremevalues for the $n=L$ case. (For clarity
            only every 40th bin is visualized, including the last bin.)
        }
    \end{figure}

    The second class of sequences $S$ we scrutinized are random walks.
    Here no analytical results are known to us, thus the
    distribution beyond the high-probability peak region seems to be
    unknown. Again, by applying the large-deviation approach,
    we sample basically the whole distribution, and
    can even compare the right tail of our distribution with the corner case of
    $L=n$, which only occurs if all increments
    $\delta$ are positive and therefore with probability $2^{-n}$. This case
    is marked in Fig.~\ref{fig:verteilungRW} to emphasize the quality of our
    data. For the left tail, we can not sample so far, as the very steep
    decline of the distribution is difficult to handle for our sampling scheme.

\begin{figure}
        \includegraphics[scale=1]{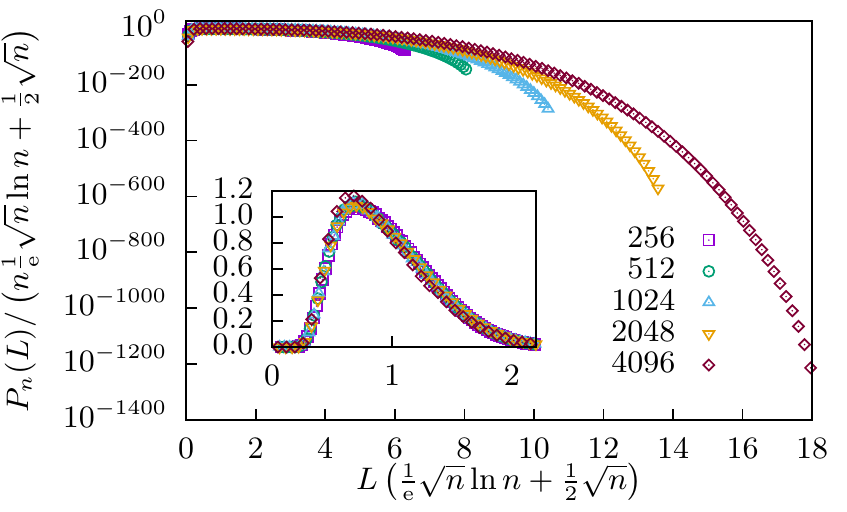}
        \caption{\label{fig:finite_rw}
            Collapse of different system sizes on a common curve $g$ from
            Eq.~\eqref{eq:rw:scaling}. Apparently the far tail shows severe
            finite-size effects, though for increasing sizes $n$ a convergence
            to a common curve is visible. (For clarity not all datapoints are drawn.)
        }
    \end{figure}

    For random walks with increments from a symmetric uniform distribution,
    indeed for increments from any symmetric distribution with finite variance,
    the scaling of the mean as $\avg{L} \propto n^\theta$ and the variance as
    $\sigma^2 \propto n^{2\theta}$ was observed in Ref.~\cite{mendoncca2017empirical}
    with $\theta = 0.5680(15)$. More interestingly the same reference suggests
    that the whole distribution follows the scaling form
    \begin{align}
        \label{eq:rw:scaling}
        P_n(L) = n^{-\theta} g(n^{-\theta} L),
    \end{align}
    with a not-explicitly known function $g$. Note the similarity to the scaling
    for the random permutation case in Eq.~\ref{eq:tw}, though here we do not
    need to subtract the mean value before rescaling with the standard deviation,
    since both happen to have the same exponent $\theta$ in this case.
    Using our data for the tails of the distribution, we can test whether this
    scaling holds over the whole distribution or only in the main region.
    If we rescale the axis of the plot suitably, the distributions for
    different sizes $n$ should collapse on the scaling function $g$,
    in the case that Eq.~\eqref{eq:rw:scaling} holds.
    Since we only have results for comparatively small systems sizes in
    comparison to Ref.~\cite{mendoncca2017empirical}, we have to include the
    logarithmic corrections to scaling, which are estimated by
    Ref.~\cite{mendoncca2017empirical} to be
    \begin{align}
        \label{eq:rw:scaling_better}
        \avg{L} \approx \frac{1}{\mathrm{e}}\sqrt{n} \ln n + \frac{1}{2}\sqrt{n}.
    \end{align}
    Note that this proposed scaling is proportional to $\sqrt{n}$, i.e., $\theta = 0.5$
    instead of the slightly larger value measured without accounting for
    logarithmic corrections. Also note that this follows the analytically
    expected form \cite{angel2017increasing}.
    In fact, using expression Eq.~\eqref{eq:rw:scaling_better}
    instead of $n^\theta$ in Eq.~\eqref{eq:rw:scaling}, leads to a better
    collapse of our data. This is shown in Fig.~\ref{fig:finite_rw},
    where the collapse does seem to work except for the very far tails, which
    is an effect -- at least partially -- caused by finite-size effects, since
    the length of the LIS can for finite $n$ never be longer than $n$. This
    pattern occurs often when looking at the far tails of
    discrete systems, e.g., for the convex hull of random walks on lattices in
    \cite{Claussen2015Convex,schawe2017highdim,schawe2018avoiding,schawe2018large}
    or in a toy model for non interacting Fermions in a landscape with $n$
    random energy levels \cite{schawe2018ground}.

\begin{figure}
        \includegraphics[scale=1]{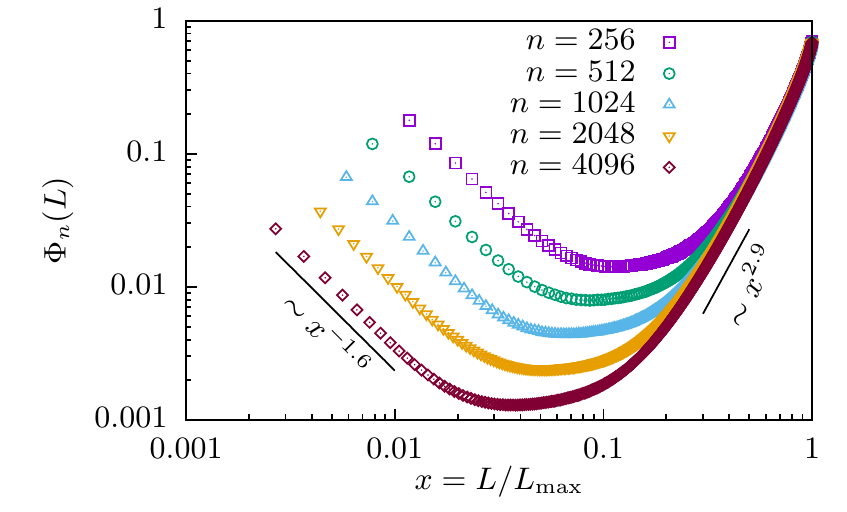}
        \caption{\label{fig:rate_rw}
            Empirical rate function $\Phi_n(L)$ for the length of the LIS of
            random walks. (For clarity not every datapoint is shown.)
        }
    \end{figure}

    Since for the rate functions characterizing the distribution of the length of
    LIS of random walks there is no known result, we use our numerical data to
    give a rough estimate of the rate function. Therefore we look into
    the empirical rate function $\Phi_n(L) = \frac{1}{n} \ln P_n(L)$,
    which is plotted in Fig.~\ref{fig:rate_rw} for the data already shown in
    Fig.~\ref{fig:verteilungRW}.

    Using the empirical rate function we can obtain the asymptotics of the rate
    function from our data. Note that to estimate the right tail
    rate function we use the intermediate tail and not the far tail, which is
    bending up due to finite-size effects. Since we are only interested in the
    leading order exponent of the rate function, i.e., assuming
    $\Phi(L) \propto L^{\kappa}$ for very small and very large values of $L$, we
    can rescale the axes arbitrarily due to the scale invariance of power laws.
    For convenience we look at $x=L/L_\mathrm{max}$ to limit the range to the
    interval $[0,1]$.
    For the left tail we observe a leading order behavior of the rate function
    of approximately $\Phi(L) \sim L^{-1.6}$ and for the right tail
    $\Phi(L) \sim L^{2.9}$, though larger values of the exponent are possible,
    if looking at a range farther right, which is probably caused by
    finite-size effects, as the very long LIS are suppressed by the hard limit
    of $L \le n$. Note that the exponent of the left tail is clearly distinct
    from $3$, such that it does not show signs of a third order phase
    transition. Also it does not show a Tracy-Widom distribution in
    the main region (also see \cite{mendoncca2017empirical}), which is
    consistent with the expectation that these two properties do occur together
    \cite{Majumdar2014Top}.

\begin{figure}
        \includegraphics[scale=1]{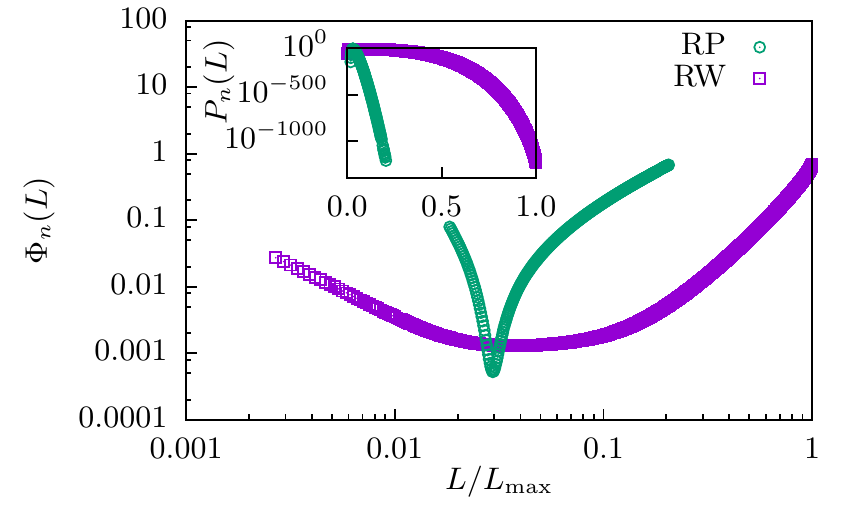}
        \caption{\label{fig:cmp_both}
            Direct comparison of the distributions for both cases, the
            random permutation and the random walk. The main plot shows the
            empirical rate function $\Phi_n(L)$. The inset shows the
            probability density $P_n(L)$.
            Both show sequences of length $n=4096$.
        }
    \end{figure}

    Comparing this leading order behavior to the behavior of the random
    permutation case as visualized in Fig.~\ref{fig:rate_rp}, shows that the tails
    decay differently. For a direct comparison of our results, consider
    Fig.~\ref{fig:cmp_both}. While the right-tail exponent is larger in the random
    walk case, the probability density decays slower (cf.~inset of
    Fig.~\ref{fig:cmp_both}). This apparent contradiction, is understandable when
    considering that the rate function of the random permutation case grows
    much faster near the minimum at $\avg{L}$, such that the rate function in
    the RP case has larger absolute values and the probability density decreases
    much faster. It is interesting that the empirical rate functions behave
    qualitatively so different for such closely related models, e.g., the
    branches left and right of the minimum show opposite curvature in the two
    cases. Generally, it is visible that the distribution $P(L)$ is much broader
    in the RW case, especially towards quite large values of $L$.

\section{Conclusions}
\label{sec:conclusion}
    We obtained numerical data for the distribution of the length of the
    longest increasing subsequence for two cases of
    sequences of random numbers, namely, for random permutations and for
    one dimensional
    random walks. By applying sophisticated large-deviation
    algorithms, we are able to sample the distributions over literally
    hundreds of decades in probability.
    The case of random permutations is already well studied in the
    analytical literature and our results confirm, to our knowledge, for the
    first time these results. Since our data is gathered for finite system sizes,
    we can observe a rather fast convergence to the analytical results valid
    in the $n\to \infty$ limit. These results show also the validity
    and convergence of our simulations.
    For the case of random walks we can observe the leading order behavior
    of the rate function far into the tails. This result could be used to guide
    analytical work on this topic or at least to test future analytical results.
    A direct comparison of the empirical rate functions in the tails shows
    qualitatively very different behavior. While the rate function of the
    random walk seems to be a convex function, the random permutation case
    consists in principle of two concave parts.

    A possible future direction extending this work would be an interpolation
    between the random permutation and random walk case, where one could observe
    the change of the exponents governing the rate function. Since a set of
    distinct random numbers $\delta_j$ drawn uniformly, from $[-1,1]$ should
    show the same statistics for the longest increasing subsequence of a random
    permutation, we could introduce a parameter $c$ governing the correlation
    length. The sequence would be
    constructed as $S_i = \sum_{j=\max(0,i-c)}^{i} \delta_j$.
    For $c=0$ this would correspond to a random permutation and for $c=n$ to
    a random walk. In addition to this simple type of correlation,
    one could study power-law correlated random numbers or increments,
    leading possibly to even more complicated behavior.

\section*{Acknowledgments}
    We are indebted to Satya N. Majumdar who brought this problem to our
    attention and gave valuable feedback on a draft of this manuscript. Also we
    want to thank J. Ricardo G. Mendon\c{c}a for interesting discussions about the
    problem and Christoph Norrenbrock for his advise during the preparation of
    the manuscript.
    We acknowledge the HPC facilities of the GWDG G\"ottingen and the CARL
    cluster in Oldenburg funded by the DFG (INST 184/157-1 FUGG) and the
    Ministry of Science and Culture (MWK) of the Lower Saxony State.
    HS acknowledges support by DFG grant HA 3169/8-1.
\bibliography{lit}

\end{document}